\documentclass[12pt]{article}
\usepackage{graphicx,psfrag,epsf,amsthm,amssymb,amsmath}
\usepackage{enumerate,natbib,dsfont,setspace,alltt}
\usepackage[ruled,vlined]{algorithm2e}
\usepackage{comment}
\usepackage{bbm}
\usepackage{url}
\usepackage{bold-extra}
\usepackage{colonequals}

\usepackage[dvipsnames,svgnames,x11names,hyperref]{xcolor}
\definecolor{americanrose}{rgb}{0.9,0.01,0.15}
\definecolor{frenchblue}{rgb}{0.01,0.15,0.9}
\definecolor{darkgreen}{rgb}{0.1,0.45,0.4}
\usepackage[colorlinks,citecolor=darkgreen,linkcolor=americanrose,
            urlcolor=frenchblue]{hyperref}

\newcommand{\blind}{1}

\def\spacingset#1{\renewcommand{\baselinestretch}%
{#1}\small\normalsize} \spacingset{1}

\newcommand{\norm}[1]{\lVert #1\rVert}
\DeclareMathOperator*{\argmin}{arg\,min}

\newtheorem{proposition}{Proposition}[section]

\newtheorem{lemma}[proposition]{Lemma}
\newtheorem{definition}[proposition]{Definition}

\def\G{\mathcal{G}}
\def\R{\mathcal{R}}
\def\U{\mathcal{U}}
\def\SU{\mathcal{SU}}
\def\SN{\mathcal{SN}}
\def\mixsqp{{\sc mix-SQP}}

\spacingset{1.05}

\begin{document}

\if1\blind { \title{\bf \bfseries{A Fast Algorithm for Maximum
      Likelihood Estimation of Mixture Proportions Using Sequential
      Quadratic Programming}}
  
  \author{Youngseok Kim\footnote{Department of Statistics, University of Chicago}$\ $,$\ $ Peter Carbonetto\footnote{Research Computing Center, University of Chicago}$^{\dagger\ddagger}$, \\
  Matthew Stephens$^*$\footnote{Department of Human Genetics, University of Chicago}$\ $,$\ $ Mihai Anitescu$^*$\footnote{Argonne National Laboratory,  corresponding author (\texttt{anitescu@anl.gov}).}}
  \maketitle
} \fi

\if0\blind
{
  \begin{center}
    {\LARGE\bf Title}
\end{center}
  \medskip
} \fi
\vspace{-2em}
\begin{abstract}
\noindent Maximum likelihood estimation of mixture proportions has a
long history, and continues to play an important role in modern
statistics, including in development of nonparametric empirical Bayes
methods. Maximum likelihood of mixture proportions has traditionally
been solved using the expectation maximization (EM) algorithm, but
recent work by Koenker \& Mizera shows that modern convex optimization
techniques---in particular, interior point methods---are substantially
faster and more accurate than EM. Here, we develop a new solution
based on sequential quadratic programming (SQP). It is substantially
faster than the interior point method, and just as accurate. Our
approach combines several ideas: first, it solves a reformulation of
the original problem; second, it uses an SQP approach to make the best
use of the expensive gradient and Hessian computations; third, the SQP
iterations are implemented using an active set method to exploit the
sparse nature of the quadratic subproblems; fourth, it uses accurate
low-rank approximations for more efficient gradient and Hessian
computations. We illustrate the benefits of the SQP approach in
experiments on synthetic data sets and a large genetic association
data set. In large data sets ($n \approx 10^6$ observations, $m
\approx 10^3$ mixture components), our implementation achieves at
least 100-fold reduction in runtime compared with a state-of-the-art
interior point solver. Our methods are implemented in Julia and in an
R package available on CRAN
(\url{https://CRAN.R-project.org/package=mixsqp}).
\end{abstract}

\medskip
\noindent%

\newpage

\section{Introduction}
\label{sec:introduction}

We consider maximum likelihood estimation of the mixture proportions
in a finite mixture model where the component densities are known. The
simplest example of this arises when we have independent and
identically distributed ({\em i.i.d.}) observations $z_1,\dots,z_n$
drawn from a finite mixture distribution with density
\begin{equation*}
p(\,\cdot \,|\, x) = \sum_{k=1}^m x_k g_k(\,\cdot\,),
\end{equation*}
where the component densities $g_k(\cdot)$ are known and $x = (x_1,
\dots, x_m)^T$ denotes the unknown mixture proportions (non-negative
and sum to one). Finding the maximum likelihood estimate (MLE) of
$x$ can be formulated as a constrained optimization problem:
\begin{equation}
\label{prob:mix_convex}
\begin{aligned}
\mbox{\rm minimize}   & \quad f(x) \triangleq -\frac1n \sum_{j=1}^n
                              \log \left(\sum_{k=1}^m L_{jk}x_k \right) \\
\mbox{\rm subject to} & \quad x \in \mathcal{S}^m \triangleq
  \big\{x : {\textstyle \sum_{k=1}^m x_k = 1},\ x \succeq 0\big\},
\end{aligned}
\end{equation}
where $L$ is an $n \times m$ matrix with entries $L_{jk} \triangleq
g_k(z_j) \geq 0$. This optimization problem arises in many
settings---including nonparametric empirical Bayes (EB)
computations described later---where observations are not necessarily
identically distributed. Here, we develop general methods for solving
\eqref{prob:mix_convex}.

Problem \eqref{prob:mix_convex} is a convex optimization problem and
can be solved simply using expectation maximization (EM)
\citep{dempster77}; see Appendix \ref{sec:em}. However, the
convergence of EM can be intolerably slow \citep{redner-walker-1984,
  atkinson-1992, salakhutdinov-2003, varadhan-2008}; this slow
convergence is illustrated evocatively in
\cite{koenker2014convex}.  \cite{koenker2014convex} and \cite{rebayes}
show that modern convex optimization methods can be substantially
faster and more reliable than EM. They demonstrate this by using an
interior (IP) method to solve a dual formulation of the original
problem. This method is implemented in the {\tt KWDual} function of
the R package {\tt REBayes} \citep{rebayes}, which interfaces to the
commercial interior point solver {\tt MOSEK}
\citep{andersen2000mosek}.

In this paper, we provide an even faster algorithm for this problem
based on sequential quadratic programming (SQP)
\citep{nocedal2006sequential}. The computational gains are greatest in
large data sets where the matrix $L \in \mathbb{R}^{n\times m}$ is
numerically rank-deficient. Rank deficiency can make the optimization
problem harder to solve, even if it is convex
\citep{wright1998superlinear}. As we show, a numerically
rank-deficient $L$ often occurs in the nonparametric EB problems that
are the primary focus of {\tt KWDual}. As an example of target problem
size, we consider data from a genome-wide association study with $n >
10^6$ and $m > 100$. For such data, our methods are approximately 100
times faster than {\tt KWDual} (about 10 s vs. 1,000 s). All our
methods and numerical experiments are implemented in the Julia
programming language \citep{bezanson2012julia}, and the source code is
available online at
\url{https://github.com/stephenslab/mixsqp-paper}. Many of our methods
are also implemented in an R package, {\tt mixsqp}, which is available
on CRAN \citep{R}.

\section{Motivation: nonparametric empirical Bayes}
\label{sec:npeb}

Estimation of mixture proportions is a fundamental problem in
statistics, dating back to at least \cite{pearson1894contributions}.
This problem, combined with the need to fit increasingly large data
sets, already provides strong motivation for finding efficient
scalable algorithms for solving \eqref{prob:mix_convex}. Additionally,
we are motivated by recent work on nonparametric approaches to
empirical Bayes (EB) estimation in which a finite mixture with a large
number of components is used to accurately approximate {\it
  nonparametric} families of prior distributions
\citep{koenker2014convex, ash}. Here we briefly discuss this
motivating application.

We first consider a simple EB approach to solving the ``normal
means,'' or ``Gaussian sequence,'' problem \citep{johnstone}. For
$j = 1, \dots, n$,  we observe data $z_j$ that are noisy observations of
some underlying ``true'' values, $\theta_j$, with normally distributed
errors of known variance $s_j^2$,
\begin{equation}
\label{eqn:z}
z_j \,|\, \theta_j \sim N(\theta_j, s_j^2).
\end{equation}
The EB approach to this problem assumes that $\theta_j$ are {\em
  i.i.d.} from some unknown distribution $g$,
\begin{equation}
\label{eqn:theta}
\theta_j \,|\, g \sim g, \quad g \in \G,
\end{equation}
where $\G$ is some specified class of distributions. The EB approach
estimates $g$ by maximizing the (marginal) log-likelihood, which is
equivalent to solving:
\begin{equation}
\label{prob:npmle}
\underset{g \,\in\, \mathcal{G}}{\textrm{minimize}} \;
-\frac{1}{n} \sum_{j=1}^n
  \log  \left[ \int N(z_j;\theta,s_j^2) \, g(d\theta) \right],
\end{equation}
where $N(\,\cdot\,; \mu, \sigma^2)$ denotes the normal density with
mean $\mu$ and variance $\sigma^2$. After estimating $g$ by solving
\eqref{prob:npmle}, posterior statistics are computed for each
$j$. Our focus is on the maximization step.

Although one can use simple parametric families for $\G$, in many
settings one might prefer to use a more flexible nonparametric
family. Examples include:
\begin{itemize}

\item $\G = \R$, the set of all real-valued distributions.

\item $\G=\U_0$, the set of unimodal distributions with a mode at
  zero. (Extensions to a nonzero mode are straightforward.)
  
\item $\G=\SU_0$, the set of symmetric unimodal distributions with a
  mode at zero.
  
\item $\G=\SN_0$, the set of distributions that are scale mixtures of
  zero-mean normals, which includes several commonly used
  distributions, such as the $t$ and double-exponential (Laplace)
  distributions.
  
\end{itemize}
The fully nonparametric case, $\G = \R$, is well studied (e.g.,
\citealt{laird1978nonparametric, jiangzhang, brown,
  koenker2014convex}) and it is related to the classic Kiefer-Wolfowitz
problem \citep{kiefer1956consistency}. More constrained examples $\G =
\U_0, \SU_0, \SN_0$ appear in \cite{ash} (see also
\citealt{cordy1997deconvolution}), and can be motivated by the desire
to shrink estimates towards zero, or to impose some regularity on $g$
without making strong parametric assumptions. For other motivating
examples, see the nonparametric approaches to the ``compound decision
problem'' described in \cite{jiangzhang} and \cite{koenker2014convex}.

The connection with \eqref{prob:mix_convex} is that these
nonparametric sets can be accurately approximated by
a finite mixture with sufficiently large number of components;
that is, they can be approximated by
\begin{equation}
\label{eqn:definition_of_g}
\G \triangleq \big\{g = {\textstyle \sum_{k=1}^m x_k g_k :
  x \in \mathcal{S}^m} \big\},
\end{equation}
for some choice of distributions $g_k$, $k=1,\ldots,m$. The $g_k$'s
are often called \emph{dictionary functions} \citep{aharon2006rm}. For
example:
\begin{itemize}
  
\item $\G=\R$: $g_k = \delta_{\mu_k}$, where $\delta_\mu$ denotes a
  delta-Dirac point mass at $\mu$, and $\mu_1,\dots,\mu_m \in
  \mathbb{R}$ is a suitably fine grid of values across the real line.
  
\item $\G=\U_0$, $\G=\SU_0$: $g_k = \text{Unif}[0,a_k],
  \text{Unif}[-a_k,0]$ or $\text{Unif}[-a_k, a_k]$, where $a_1,
  \ldots, a_m \in \mathbb{R}^{+}$ is a suitably large and fine grid of
  values.
  
\item $\G=\SN_0$: $g_k = N(0,\sigma^2_k)$, where $\sigma_1^2, \ldots,
  \sigma_m^2 \in \mathbb{R}^{+}$ is a suitably large and fine grid of
  values.
  
\end{itemize}
With these approximations, solving \eqref{prob:npmle} reduces to
solving an optimization problem of the form \eqref{prob:mix_convex},
with $L_{jk} = \int N(z_j; \theta,s_j^2) \, g_k(d\theta)$,
the convolution of $g_k$ with a normal density $N(z_j; \theta, s_j^2)$.

A common feature of these examples is that they all use a fine grid to
approximate a nonparametric family. The result is that {\em many of
  the distributions $g_k$ are similar to one another.} Hence, the
matrix $L$ is numerically rank deficient and, in our experience, many
of its singular values are near floating-point machine precision. We
pay particular attention to this property when designing our
optimization methods.

The normal means problem is just one example of a broader class of
problems with similar features. The general point is that
nonparametric problems can often be accurately solved with finite
mixtures, resulting in optimization problems of the form
\eqref{prob:mix_convex}, typically with moderately large $m$, large
$n$, and a numerically rank-deficient $n \times m$ matrix $L$.

\section{A new SQP approach}
\label{sec:convex_programming}

The methods from \cite{rebayes}, which are implemented in function
{\tt KWDual} from R package {\tt REBayes} \citep{rebayes} provide, to our
knowledge, the best current implementation for solving
\eqref{prob:mix_convex}. These methods are based on reformulating
\eqref{prob:mix_convex} as
\begin{equation}
\label{prob:npmle_sub}
\underset{x\,\in\,\mathcal{S}^m}{\textrm{minimize}}
\quad -\frac1n \sum_{j=1}^n \log y_j \quad
\textrm{subject to} \quad Lx = y,
\end{equation}
then solving the dual problem (``K-W dual'' in \citealt{koenker2014convex}),
\begin{equation}
\label{prob:npmle_dual}
\underset{\nu\,\in\,\mathbb{R}^m}{\textrm{minimize}} \quad
-\frac1n \sum_{j=1}^n \log \nu_j \quad
\textrm{subject to} \quad L^T \nu \preceq n \mathds{1}_m, \quad
\nu \succeq 0, 
\end{equation}
where $\mathds{1}_m$ is a vector of ones of length $m$.
\cite{koenker2014convex} reported that solving this dual formulation was
generally faster than primal formulations in their assessments.
Indeed, we also found this to be the case for IP approaches (see
Figure \ref{fig:primal_dual}). For $n\gg m$, however, we believed that
the original formulation \eqref{prob:mix_convex} offered more
potential for improvement. In the dual formulation
\eqref{prob:npmle_dual}, effort depends on $n$ when computing the
gradient and Hessian of the objective, when evaluating the
constraints, and when computing the Newton step inside the IP
algorithm. By contrast, in the primal formulation
\eqref{prob:mix_convex}, effort depends on $n$ only in the gradient and
Hessian computations; all other evaluations depend on $m$ only.

These considerations motivated the design of our algorithm. It was
developed with two key principles in mind: (i) make best possible use
of each expensive gradient and Hessian computation in order to
minimize the number of gradient and Hessian evaluations; and (ii)
reduce the expense of each gradient and Hessian evaluation as much as
possible. (We could have avoided Hessian computations by pursuing a
first-order optimization method, but we judged that a second-order
method would likely be more robust and stable because of the
ill-conditioning caused by the numerical rank deficiency of $L$; we briefly
investigate the potential of first-order optimization methods in
Section \ref{sec:mixsqp-vs-first-order}.)

To make effective use of each Hessian computation, we apply sequential
quadratic programming  \citep{nocedal2006sequential} to a
reformulation of the primal problem \eqref{prob:mix_convex}. SQP
attempts to make best use of the expensive Hessian computations by
finding, at each iteration, the best reduction in a quadratic
approximation to the constrained optimization problem. 

To reduce the computational cost of each Hessian evaluation, we use a
``rank-revealing'' QR decomposition of $L$ to exploit the numerical
low-rank of $L$ \citep{golub2012matrix}. The RRQR matrix
decomposition, which need be performed only once, reduces
subsequent per-iteration computations so that they depend on the
numerical rank, $r$, rather than $m$. In particular, Hessian computations
are reduced from $O(nm^2)$ to $O(nr^2)$.

In addition to these two key principles, two other design decisions
were also important to reduce effort. First, we introduced a
reformulation that relaxes the simplex constraint to a less
restrictive non-negative one, which simplifies computations. Second,
based on initial observations that the primal solution is often
sparse, we implemented an active set method
\citep{nocedal2006sequential}---one that estimates which entries of the
solution are zero (this is the ``active set'')---to solve for the
search direction at each iteration of the SQP algorithm. As we show
later, an active set approach effectively exploits the solution's
sparsity.

The remaining subsections detail these innovations.

\subsection{A reformulation}
\label{sec:reform}

We transform \eqref{prob:mix_convex} into a simpler problem with less
restrictive non-negative constraints using the following definition
and proposition.
\begin{definition}
\label{def:scaleinvariant}
  A function $\phi: \mathbb{R}_{+}^{m} \rightarrow \mathbb{R}$ is said to
be ``scale invariant'' if for any $c > 0$ there exists a $C \in
\mathbb{R}$ such that for any $x \in \mathbb{R}_{+}^{m}$ we have
$\phi(cx) = \phi(x) - C$.
\end{definition}

\begin{proposition}
\label{thm:equivalence}
Consider the simplex-constrained optimization problem 
\begin{equation}
\label{prob:simplex}
\underset{x\,\in\,\mathcal{S}^m}{\rm minimize} \; \phi(x),
\end{equation}
where $\phi(x)$ is scale invariant, convex, and nonincreasing with
respect to $x$---that is, $x \succeq y$ (the componentwise partial
ordering) implies $\phi(x) \leq \phi(y)$ for all $x \in
\mathbb{R}_{+}^{m}$. Let $x^{*}(\lambda)$ denote the solution to a
Lagrangian relaxation of \eqref{prob:simplex},
\begin{equation}
\label{prob:reformulated_simplex}
\underset{x\,\in\,\mathbb{R}^m}{\rm minimize} \quad
\phi_{\lambda}(x) \triangleq \phi(x) +
  \lambda \sum_{k=1}^m x_k, \quad \mbox{\rm subject to} \quad
  x \succeq 0,
\end{equation}
for $\lambda > 0$. Then $x^{*} \triangleq x^{*}(\lambda) /
\sum_{k=1}^m x_k^{*}(\lambda)$ is a solution to
\eqref{prob:simplex}.
\end{proposition}

By setting $\phi$ to $f$, the objective function in
\eqref{prob:mix_convex}, this proposition implies that
\eqref{prob:mix_convex} can be solved by instead solving
\eqref{prob:reformulated_simplex} for some $\lambda$. Somewhat
surprisingly, in this special case setting $\lambda=1$ yields a
solution $x^*(\lambda)$ that is already normalized to sum to 1, so
$x^* = x^*(1)$. This result is summarized in the following
proposition:

\begin{proposition}
\label{corr:special_equiv}
Solving the target optimization problem \eqref{prob:mix_convex} is
equivalent to solving \eqref{prob:reformulated_simplex} with
$\phi = f$ and $\lambda=1$; that is,
\begin{equation}
\label{prob:main_problem}
\underset{x\,\in\,\mathbb{R}^m}{\rm minimize} \quad
f^{\star}(x) \triangleq f(x) + \sum_{k=1}^m x_k, \quad
\mbox{\rm subject to} \quad x \succeq 0.
\end{equation}
\end{proposition}
\noindent The proofs of the propositions are given in the
Appendix.

While we focus on the case of $\phi = f$, these ideas should apply
to other objective functions so long as they satisfy
Definition~\ref{def:scaleinvariant}; e.g., when $f$ is a composite of
``easily differentiable'' scale-invariant functions and ``thin and
tall'' linear functions. Many of the algorithmic ideas presented in
following sections are applicable to those functions as well. See
the Appendix for further discussion.

\subsection{Sequential quadratic programming}
\label{ss:SQP}

We solve the reformulated optimization problem
\eqref{prob:main_problem} using an SQP algorithm with backtracking
line search \citep{nocedal2006sequential}. In brief, SQP is an
iterative algorithm that, at the $t$-th iteration, computes a
second-order approximation of the objective at the feasible point
$x^{(t)}$, then determines a search direction $p^{(t)}$ based on the
second-order approximation. At iteration $t$, the search direction
$p^{(t)}$ is the solution to the following non-negatively-constrained
quadratic program:
\begin{equation}
\label{prob:sqp_subproblem}
p^{(t)} = \argmin_{p\,\in\,\mathbb{R}^m} \textstyle
\frac{1}{2} p^TH_tp + p^Tg_t \quad
\textrm{subject to} \quad x^{(t)} + p \succeq 0,
\end{equation}
where $g_t = \nabla f^{\star}(x^{(t)})$ and $H_t = \nabla^2
f^{\star}(x^{(t)})$ are the gradient and Hessian of $f^{\star}(x)$ at
$x^{(t)}$. Henceforth, this is called the ``quadratic subproblem.''
Computation of the gradient and Hessian is considered in Section
\ref{ss:gradient}, and solving the quadratic subproblem is discussed
in Section \ref{subsec:QP}.

After identifying the search direction, $p^{(t)}$, the SQP method
performs a backtracking line search to determine a sufficient descent
step $x^{(t+1)} = x^{(t)} + \alpha_t p^{(t)}$, for $\alpha_t \in
(0,1]$. In contrast to other projection-based methods such as the
projected Newton method \citep{kim2010tackling}, $x^{(t+1)}$ is
guaranteed to be (primal) feasible for all choices of $\alpha_t \in
(0,1]$ provided that $x^{(t)}$ is feasible. This is due to the
linearity of the inequality constraints. As discussed in
\cite{nocedal2006sequential}, the line search will accept unitary
steps ($\alpha_t = 1$) close to the solution and the iterates will
achieve quadratic convergence provided the reduced Hessian is positive
definite for all co-ordinates outside the optimal active set (the
indices that are zero at the solution $x^{*}$). A similar result
can be found in \cite{wright1998superlinear}.

\subsection{Gradient and Hessian evaluations}
\label{ss:gradient}

We now discuss computation of the gradient and Hessian and ways to
reduce the burden of computing them.
\begin{lemma}
\label{lemma:structured}
For any $x \in \mathbb{R}_+^m$, the gradient and Hessian of the objective
function in \eqref{prob:main_problem} are given by
\begin{equation}
\label{eqn:grad_hess}
\textstyle
g = \nabla f^{\star}(x) = -\frac{1}{n} L^T d + \mathds{1}_m,\quad
H = \nabla^2 f^{\star}(x) = \frac{1}{n} L^T {\rm diag}(d)^2 L,
\end{equation}
where $\mathds{1}_m$ is a vector of ones of length $m$, and $d = (d_1,
\ldots, d_n)^T$ is a column vector with entries $d_j =
1/(Lx)_j$. Further, for any $x \in \mathbb{R}_+^m$ the following
identities hold:
\begin{equation}
\label{eqn:structured_grad_hess}
x^Tg = 0,\quad x^THx = 1,\quad Hx + g = \mathds{1}_m.
\end{equation}
\end{lemma}

\noindent Assuming $L$ is not sparse, computing $g$ (and $d$) requires
$O(nm)$ multiplications, and computing $H$ requires $O(nm^2)$
multiplications. The result \eqref{eqn:structured_grad_hess} is easily
derived by substitution from \eqref{eqn:grad_hess}.

In practice, we find that $L$ is often numerically rank deficient,
with (numerical) rank $r < m$. We can exploit this property to
reduce computational effort by approximating $L$ with a low-rank
matrix.  We use either the RRQR decomposition \citep{golub2012matrix}
or a truncated singular value decomposition (tSVD) to compute an accurate,
low-rank approximation to $L$. Specifically, we use the {\tt pqrfact}
and {\tt psvdfact} functions from the {\tt LowRankApprox} Julia
package \citep{lowrankapprox} which implement randomized algorithms
based on \cite{halko2011finding}.

The rank-$r$ QR approximation of $L$ is $\tilde{L} = QRP^T$, with $Q
\in \mathbb{R}^{n\times r}$, $R \in \mathbb{R}^{r\times m}$ and $P \in
\mathbb{R}^{m\times m}$, the permutation matrix. Therefore, a rank-$r$
QR decomposition yields an approximate gradient and Hessian:
\begin{equation}
\label{eqn:grad_hess_qr}
\tilde{g} = \textstyle -\frac{1}{n} PR^TQ^T\tilde{d} + \mathds{1}_m, \quad
\tilde{H} = \textstyle \frac{1}{n} PR^TQ^T \textrm{diag}(\tilde{d})^2 QRP^T,
\end{equation}
where $\tilde{d}$ is a vector of length $n$ with entries $\tilde{d}_j
= 1/(QRP^Tx)_j$. Corresponding expressions for the tSVD are
straightforward to obtain, and are therefore omitted.

Once we have obtained a truncated QR (or SVD) approximation to $L$,
the key to reducing the expense of the gradient and Hessian
computations is to avoid directly reconstructing $\tilde{L}$. For
example, computing $\tilde{g}$ is implemented as $\tilde{g} =
-(((\tilde{d}^TQ)R)P^T)^T/n + \mathds{1}_m$. In this way, all matrix
multiplications are a matrix times a vector. The dominant cost in
computing $\tilde{H}$ is the product $Q^T \textrm{diag}(\tilde{d})^2 Q$, which
requires $O(nr^2)$ multiplications. Overall, computation is reduced by
roughly a factor of $(r/m)^2$ per iteration compared with
\eqref{eqn:grad_hess}.  To enjoy this benefit, we pay the one-time
cost of factorizing $L$, which, in the regime $n \gg m$, is $O(nmr)$
\citep{golub2012matrix}.

\subsection{Solving the quadratic subproblem}
\label{subsec:QP}

To find the solution to the quadratic subproblem
\eqref{prob:sqp_subproblem}, we set $p^{*} = y^{*} - x^{(t)}$ in which
$y^{*}$ is the solution to
\begin{equation}
\label{prob:sqp_subproblem_reform}
\underset{y\,\in\,\mathbb{R}^m}{\textrm{minimize}} \;
\textstyle \frac{1}{2} y^T H_t y + y^T a_t, \quad
\textrm{subject to} \quad y \succeq 0
\end{equation}
with $a_t = -H_t x^{(t)} + g_t = 2g_t - \mathds{1}_m$. This problem
comes from substituting $y = x^{(t)} + p$ into
\eqref{prob:sqp_subproblem}. Solving
\eqref{prob:sqp_subproblem_reform} is easier than
\eqref{prob:sqp_subproblem} due to the simpler form of the inequality
constraints.

To solve \eqref{prob:sqp_subproblem_reform}, we implement an active
set method following \citet[\S 16.5]{nocedal2006sequential}. The
active set procedure begins at a feasible point $y^{(0)}$ and an
initial estimate of the active set, $\mathcal{W}^{(0)}$ (the ``working
set''), and stops when the iterates $y^{(0)}, y^{(1)}, y^{(2)},
\ldots$ converge to a fixed point of the inequality-constrained
quadratic subproblem \eqref{prob:sqp_subproblem_reform}. The initial
working set $\mathcal{W}^{(0)}$ and estimate $y^{(0)}$ can be set to
predetermined values, or they can be set according to the current SQP
iterate, $x^{(t)}$ (this is often called ``warm-starting''). We
initialized the active set solver to $x^{(0)} = \{\frac{1}{m}, \ldots,
\frac{1}{m}\}$ in the first iteration of SQP, and used a warm start in
subsequent iterations.

The active set method is an iterative method in which the $l$th
iteration involves solving an equality-constrained problem,
\begin{equation} 
\label{prob:active_set_subprob}
\underset{q \,\in\, \mathbb{R}^m}{\textrm{minimize}} \; \textstyle
\frac{1}{2} q^T H_t q + q^T b_l \quad\textrm{subject to} \quad q_i = 0,
\; \forall i \in \mathcal{W}^{(l)},
\end{equation} 
where $b_l = H_t y^{(l)} + a_t$. The solution to
\eqref{prob:active_set_subprob} at the $l$th iteration, $q^{(l)}$,
yields a search direction for the next iterate, $y^{(l+1)}$. Computing
the solution to \eqref{prob:active_set_subprob} reduces to solving a
system of linear equations, with one equation for each co-ordinate
outside the working set. Therefore, if the number of inactive
co-ordinates remains much smaller than $m$, we expect the complexity
of the active set step to be much smaller than $O(m^3)$.

Additional details of the active set implementation, including the
updates to the working set in the presence of blocking constraints,
are given in the Appendix.

\subsection{The ``\bfseries{\scshape{mix-SQP}}'' algorithm}
\label{ss:mixSQP}

Putting these components together results in Algorithm \ref{alg:sqp},
which we call \mixsqp{}. We give the algorithm for the case when RRQR
is used to approximate $L$; variants of \mixsqp{} using the truncated
SVD, or with no approximation to $L$, are similar. The most
complicated part to implement is the active set method; the details of
this step are given separately in Algorithm \ref{alg:subprob} in the
Appendix.

\begin{algorithm}[tb]
\SetKwInOut{Input}{Inputs}
\SetKwInOut{Output}{Output}
\Input{likelihood matrix, $L \in \mathbb{R}^{n\times m}$; initial
  estimate, $x^{(0)} \in \mathcal{S}_m$; stringency of sufficient
  decrease condition, $0 < \xi <1$ (default is $0.01$); step size
  reduction in backtracking line search, $0 < \rho < 1$ (default is
  $0.5$); SQP convergence tolerance, $\epsilon_{\mathrm{dual}} \geq 0$
  (default is $10^{-8}$); convergence tolerance for active set step,
  $\epsilon_{\mbox{\scriptsize active-set}} \geq 0$ (default is $10^{-10}$).}
\Output{$x^{(t)} \in \mathbb{R}^m$, an estimate of the solution to
  \eqref{prob:mix_convex}.}
Compute RRQR factorization $\{Q,R,P\}$ of $L$; \\
\For {t {\rm = 0, 1, 2, \ldots}} {
  Compute approximate gradient $\tilde{g}_t$ and Hessian $\tilde{H}_t$
  at $x^{(t)}$; (see eq.~\ref{eqn:grad_hess_qr}) \\
  $\mathcal{W} \leftarrow \{1, \ldots, m\} \setminus \mathrm{supp}(x^{(t)})$;
  \, (current estimate of working set) \\
  $y \leftarrow \mbox{\sc mix-active-set}(\tilde{g}_t, \tilde{H}_t,
  \mathcal{W}, \epsilon_{\mbox{\scriptsize active-set}})$;
  \, (see Algorithm \ref{alg:subprob}) \\
  $p^{(t)} \leftarrow y - x^{(t)}$; \\
\If{$\min_k(\tilde{g}_t)_k \geq -\epsilon_{\mathrm{dual}}$} {
{\bf stop}; \, (maximum dual residual of KKT conditions is less than
  $\epsilon_{\mathrm{dual}}$)
}
$\alpha_t \leftarrow 1$; \, (backtracking line search) \\
\While{$\tilde{f}(x^{(t)} + \alpha_t p^{(t)}) >
  \tilde{f}(x^{(t)}) + \xi \alpha_t \tilde{g}_t^T p^{(t)}$} {
  $\alpha_t \leftarrow \rho \alpha_t$;
}
$x^{(t+1)} \leftarrow x^{(t)} + \alpha_t p^{(t)}$;
}
\caption{\mixsqp{} with RRQR approximation of $L$.}
\label{alg:sqp}
\end{algorithm}

\subsection{Practical implementation details} 

A useful property of problem \eqref{prob:mix_convex} is that the
gradient \eqref{eqn:grad_hess} is unaffected by the ``scale'' of the
problem; for example, if we multiply all entries of $L$ by 100, the
gradient remains the same. This property has several practical
benefits; for example, the ``dual residual,'' used to assess
convergence of the iterates, is invariant to the scale of $L$.

When we replace $L$ with an approximation, for example 
 $\tilde{L} = QRP^T$, we are effectively solving an approximation to
\eqref{prob:main_problem},
\begin{equation}
\label{eqn:QR_approx_problem}
\underset{x\,\in\,\mathbb{R}^m}{\textrm{minimize}} \quad
\tilde{f}(x) \triangleq -\frac{1}{n}
\sum_{j=1}^n \log \left(\sum_{k=1}^m \tilde{L}_{jk}x_k \right) +
\sum_{k=1}^m x_k \quad  \textrm{subject to} \quad x \succeq 0.
\end{equation}
When an approximated likelihood matrix $\tilde{L}$ is used, some
entries $\tilde{L}_{jk}$ may have negative values, and thus the terms
inside the logarithms, $\sum_{k=1}^m \tilde{L}_{jk}x_k$, can be
slightly below zero at some feasible points. This can occur either due
to the randomized nature of the matrix decomposition algorithms we
used, or due to the finite precision of the low-rank approximations.
This is a critical point to attend to since the logarithm in the
objective does not accept negative values, for example when the
objective of \eqref{eqn:QR_approx_problem} is evaluated during line
search. In principle, this is not a problem so long as the initial
point $x^{(0)}$ satisfies $\sum_{k=1}^m \tilde{L}_{jk}x_k^{(0)} >
0$ for all $j$. Indeed, one can refine the problem statement as
\begin{equation}
\label{prob:extended_problem}
\underset{x\,\in\,\mathbb{R}^m}{\textrm{minimize}} \quad
\tilde{f}(x) \quad \textrm{subject to}
\quad x \succeq 0, \; \tilde{L}x \succeq 0,
\end{equation}
and start with a feasible $x^{(0)}$. In practice, we implemented a
simple workaround: we added a small positive constant (typically
somewhere between $10^{-8}$ and $10^{-6}$) to all the terms inside the
logarithms so as to ensure that they were strictly bounded away from
zero for all $x \in \mathcal{S}^m$.

\section{Numerical experiments}
\label{sec:numerical}

We conducted numerical experiments to compare different methods for
solving problems of the form \eqref{prob:mix_convex}. We considered
problems of this form that arise from nonparametric EB, with $\G =
\SN_0$ (Section \ref{sec:npeb}). Our comparisons involved simulating
data sets with varying numbers of data points ($n$) and grid sizes
($m$). We also evaluated the methods on a large-scale genetic data
set.

\subsection{Data sets}

For each synthetic data set, we generated $z_1, \ldots, z_n$
independently as
\begin{equation*}
z_j \,|\, \theta_j \sim N(\theta_j,1),
\end{equation*}
where the means $\theta_j$ were {\em i.i.d.} random draws from $g$, a
heavy-tailed symmetric distribution about zero,
\begin{equation*}
g = 0.5
N(0,1) + 0.2 t_4 + 0.3 t_6.
\end{equation*}
Here, $t_{\nu}$ denotes the density of Student's $t$ distribution with
$\nu$ degrees of freedom.

In addition, we used data generated by the GIANT consortium
(``Genetic Investigation of ANthropometric Traits'') for investigating
the genetic basis of human height \citep{giant}. We used the additive
effects on height estimated for $n = \mbox{2,126,678}$ genetic
variants (single nucleotide polymorphisms, or SNPs) across the
genome. Height is a well-studied example of trait with a complex
genetic basis, so the distribution of additive effects is expected to
be representative of genetic associations for many other complex
traits and diseases \citep{gwas-catalog}. The data consist of the
estimated effect sizes, $z_j$, and their corresponding standard errors,
$s_j$. For illustration, we treat the $n$ data points as
independent, though in practice there are local correlations between
SNPs. See the Appendix for more details on steps taken to download and
prepare the GIANT data.

We modeled all data sets using the ``adaptive shrinkage''
nonparametric EB model from \citet{ash} (see Sec.~\ref{sec:npeb}). The
R package {\tt ashr}, available on CRAN \citep{ashr}, implements
various versions of adaptive shrinkage; for our experiments, we
re-coded the simplest version of adaptive shrinkage in Julia. This
assumes a normal likelihood and a prior that is a finite mixture of
zero-centered normals ($\G = \SN_0$ in Sec.~\ref{sec:npeb}). This
leads to matrix entries $L_{jk} = N(z_j; 0, \sigma_k^2 + s_j^2)$, the
normal density at $z_j$ with zero mean and variance $\sigma_k^2 +
s_j^2$, where $\sigma_k^2$ is the variance of the $k$th mixture
component. For the simulated data, we set $s_j = 1$; for the GIANT
data, we set the $s_j$'s to the standard errors provided. The adaptive
shrinkage method also requires a grid of variances, $\sigma_1^2,
\dots, \sigma_m^2$. We ran \mixsqp{} on matrices $L$ with a range of
grid sizes, $m$, and for each grid size we used the method from
\cite{ash} to select the grid values. To avoid numerical overflow or
underflow, each row of $L$ was computed up to a constant of
proportionality such that the largest entry in each row was always
1. (Recall the maximum likelihood estimate is invariant to scaling the
rows of $L$.)

\subsection{Approaches considered}

Most optimization methods tested in our experiments are combinations
of the following four elements:
\begin{enumerate}

\item {\bf Problem formulation:} The method either solves the dual
  \eqref{prob:npmle_dual}, simplex-constrained
  \eqref{prob:mix_convex}, or non-negatively-constrained formulation
  \eqref{prob:main_problem}. This choice is indicated by {\tt D}, {\tt
    S}, or {\tt NN}, respectively.

\item {\bf Solver:} The optimization problem is solved using an SQP or
  IP solver. This choice is denoted by {\tt SQP} and {\tt IP},
  respectively.

\item {\bf QP solver:} For SQP approaches only, we considered two
  ways to solve the quadratic subproblem
  \eqref{prob:sqp_subproblem}: an active set method (see 
  Section~\ref{subsec:QP}) or an off-the-shelf QP solver (the
  commercial IP solver {\tt MOSEK}). We indicate this choice with {\tt
    A} or {\tt IP}, respectively. When the SQP method is not used, we
  indicate this choice with {\tt NA}, for ``not applicable.''

\item {\bf Gradient and Hessian computation:} The objective and
  partial derivatives either are computed exactly (within
  floating-point precision) by using the full matrix $L$, or
  approximated using a truncated SVD or RRQR decomposition of $L$
  (Section \ref{ss:gradient}). We denote this choice by {\tt F} (for
  ``full matrix''), {\tt SVD} or {\tt QR}.
  
\end{enumerate}
An optimization method is therefore fully specified by
\begin{center}
{\tt [formulation]-[solver]-[QP solver]-[gradient/Hessian computation]}.
\end{center}
For example, the \mixsqp{} method with a RRQR approximation $L$
(Algorithm \ref{alg:sqp}) is written as {\tt NN-SQP-A-QR}. We also
assessed the performance of two methods that do not use second-order
information: projected gradient descent \citep{birgin-2000,
  schmidt2009} and EM (see Appendix \ref{sec:em}).

All numerical comparisons were run on machines with an Intel Xeon
E5-2680v4 (``Broadwell'') processor. Other than the projected gradient
method, all methods were run from Julia version 0.6.2
\citep{bezanson2012julia} linked to the OpenBLAS optimized numerical
libraries that were distributed with Julia. (The code has also been
updated for Julia 1.1.) The {\tt KWDual} function in R package {\tt
  REBayes} \citep{rebayes} was called in R 3.4.1 and was interfaced to
Julia using the {\tt RCall} Julia package. The KWDual function used
version 8.0 of the MOSEK optimization library. The projected gradient
method was run in MATLAB 9.5.0 (2018b) with optimized Intel MKL
numerical libraries. Julia source code and scripts implementing the
methods compared in our experiments, including Jupyter notebooks
illustrating how the methods are used, are available at
\url{https://github.com/stephenslab/mixsqp-paper}. The \mixsqp{}
method is also implemented as an R package, {\tt mixsqp}, which is
available on CRAN and GitHub
(\url{https://github.com/stephenslab/mixsqp}).

\subsection{Results}

\subsubsection{Comparison of problem formulations}

\begin{figure}[tb!]
\begin{center}
\includegraphics[width=6.25in]{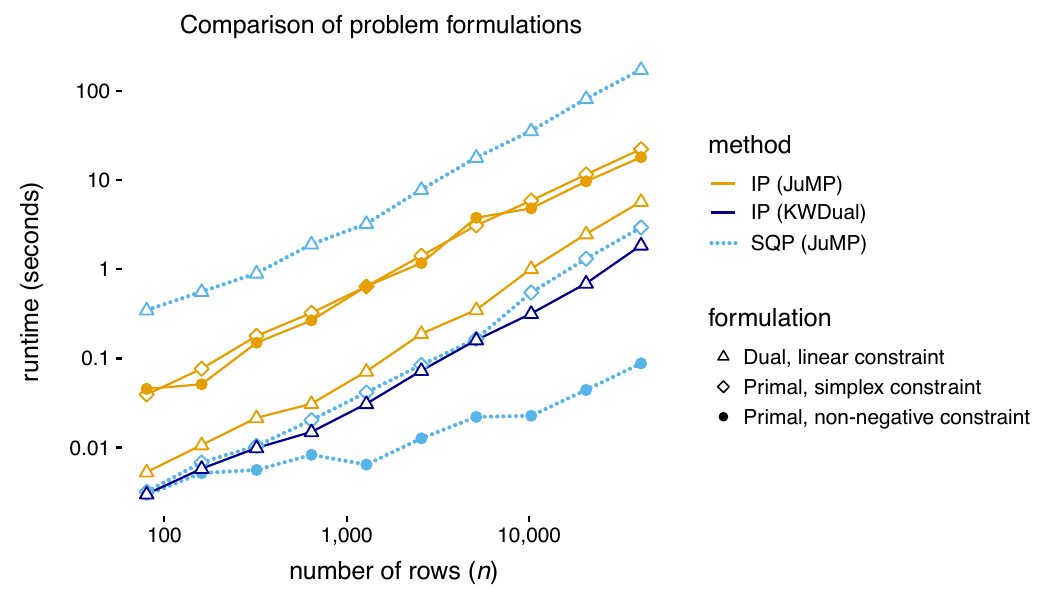} 
\caption{Runtimes for different formulations of the maximum likelihood
  estimation problem: dual \eqref{prob:npmle_dual},
  simplex-constrained \eqref{prob:mix_convex} and
  non-negatively-constrained \eqref{prob:main_problem}. For each
  problem formulation, we applied an IP or SQP algorithm. As a
  baseline, we compared against the {\tt KWDual} function from the
  {\tt REBayes} package which solves the dual formulation using {\tt
    MOSEK}. Results are from data sets with $m = 40$ and $n$ varying
  from 80 to 40,960. Runtimes are averages over 10 independent
  simulations.}
\label{fig:primal_dual}
\end{center}
\end{figure}

First, we investigated the benefits of the three problem formulations:
the dual form \eqref{prob:npmle_dual}, the simplex-constrained form
\eqref{prob:mix_convex} and the non-negatively-constrained form
\eqref{prob:main_problem}. We implemented IP and SQP methods for each
of these problem formulations in the {\tt JuMP} modeling environment
\citep{DunningHuchetteLubin2017}. We applied these methods to $n
\times m$ simulated data sets $L$, with $m = 40$ and $n$ ranging from
80 to 40,960. For all SQP solvers, an IP method was used to solve the
quadratic subproblems. In summary, we compared solvers {\tt $x$-IP-NA-F}
and {\tt $x$-SQP-IP-F}, substituting {\tt D}, {\tt S} or {\tt NN} for
$x$. In all cases, the commercial solver {\tt MOSEK} was used to
implement the IP method. To provide a benchmark for comparison, we
also ran the {\tt KWDual} method in R, which calls {\tt MOSEK}.

The results of this experiment are summarized in
Figure~\ref{fig:primal_dual}. All runtimes scaled approximately
linearly in $n$ (the slope is near 1 on the log-log scale). Of the
methods compared, SQP applied to the non-negatively-constrained
formulation, {\tt NN-SQP-IP-F}, consistently provided the fastest
solution. 

SQP for the non-negatively-constrained formulation was substantially
faster than SQP for the simplex-constrained formulation. The former
typically required fewer outer iterations, but this does not
completely explain the difference in performance---it is possible that the
simplex-constrained formulation could be improved with more
careful implementation using the {\tt JuMP} interface.

Of the IP approaches, the fastest was the implementation using the
dual formulation. This is consistent with the results of
\cite{koenker2014convex}. By contrast, the SQP approach appears to be
poorly suited to the dual formulation.

Based on these results, we focused on the non-negatively-constrained
formulation for SQP in subsequent comparisons.

\subsubsection{Examining the benefits of approximating $L$}
\label{sec:assess_approx}

\begin{figure}[tb!]
\begin{center}
\includegraphics[width=\textwidth]{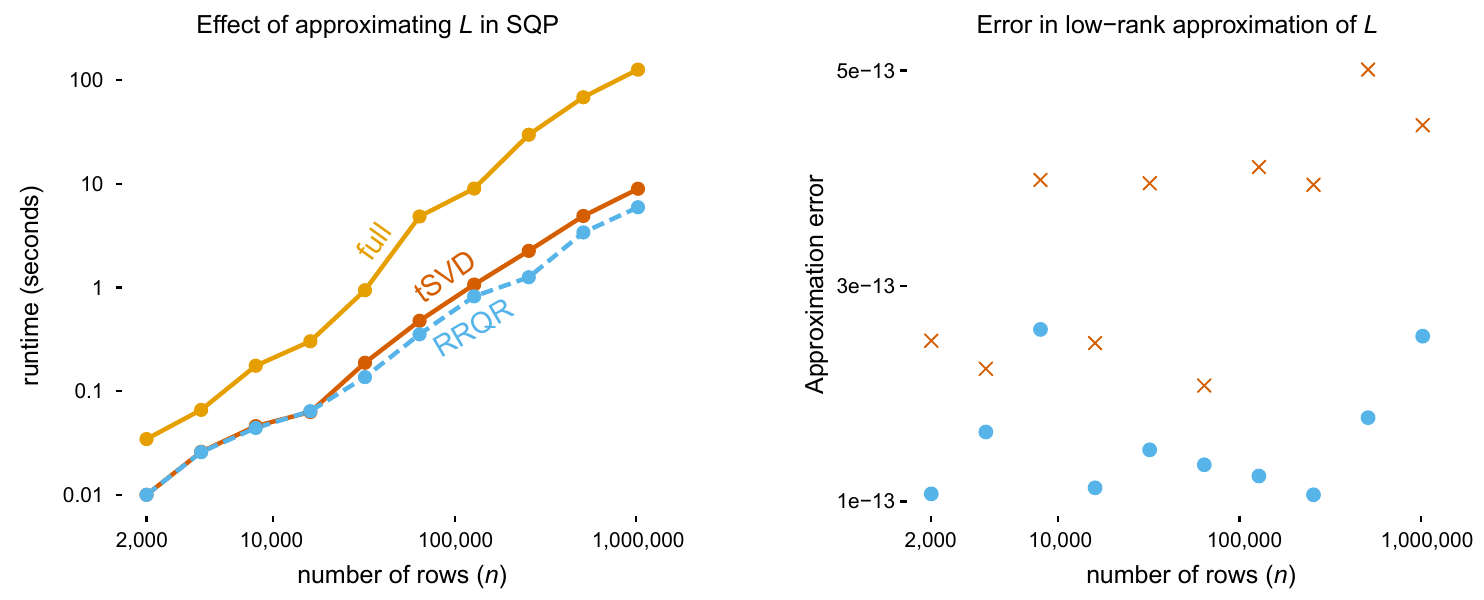}
\caption{Comparison of SQP methods with and without low-rank
  approximations to the $n \times m$ matrix, $L$. {\em Left panel:}
  Runtime of the SQP solver using the full matrix (``full'') and
  using low-rank approximations based on RRQR and tSVD
  factorizations. Results are from data sets with $m = 100$ and
  varying $n$. {\em Right panel:} Reconstruction error in RRQR (blue
  circles) and tSVD (red crosses); error is computed as $\norm{L -
    \tilde{L}}_F$. All runtimes and errors are averages over 10
  simulations.}
\label{fig:svd_qr}
\end{center}
\end{figure}

Next, we investigated the benefits of exploiting the low-rank
structure of $L$ (see Section \ref{ss:gradient}).  We compared the
runtime of the SQP method with and without low-rank approximations,
RRQR and tSVD; that is, we compared solvers {\tt NN-SQP-A-$x$}, with
$x$ being one of {\tt F}, {\tt QR}, or {\tt SVD}. We applied the three
SQP variants to $L$ matrices with varying numbers of rows. We used
functions {\tt pqrfact} and {\tt psvdfact} from the {\tt
  LowRankApprox} Julia package \citep{lowrankapprox} to compute the
RRQR and tSVD factorizations. For both factorizations, we set the
relative tolerance to $10^{-10}$.

The left-hand panel in Figure~\ref{fig:svd_qr} shows that the SQP
method with an RRQR and tSVD approximation of $L$ was consistently
faster than running SQP with the full matrix, and by a large margin;
e.g., at $n = 10^6$, the runtime was reduced by a factor of over
10. At the chosen tolerance level, these low-rank approximations
accurately reconstructed the true $L$ (Figure~\ref{fig:svd_qr},
right-hand panel).

For the largest $n$, SQP with RRQR was slightly faster than SQP with
the tSVD. We attribute this mainly to the faster computation of the
RRQR factorization. We found that the SQP method took nearly the same
solution path regardless of the low-rank approximation method used
(results not shown).

\begin{figure}[tb!]
\begin{center}
\includegraphics[width=\textwidth]{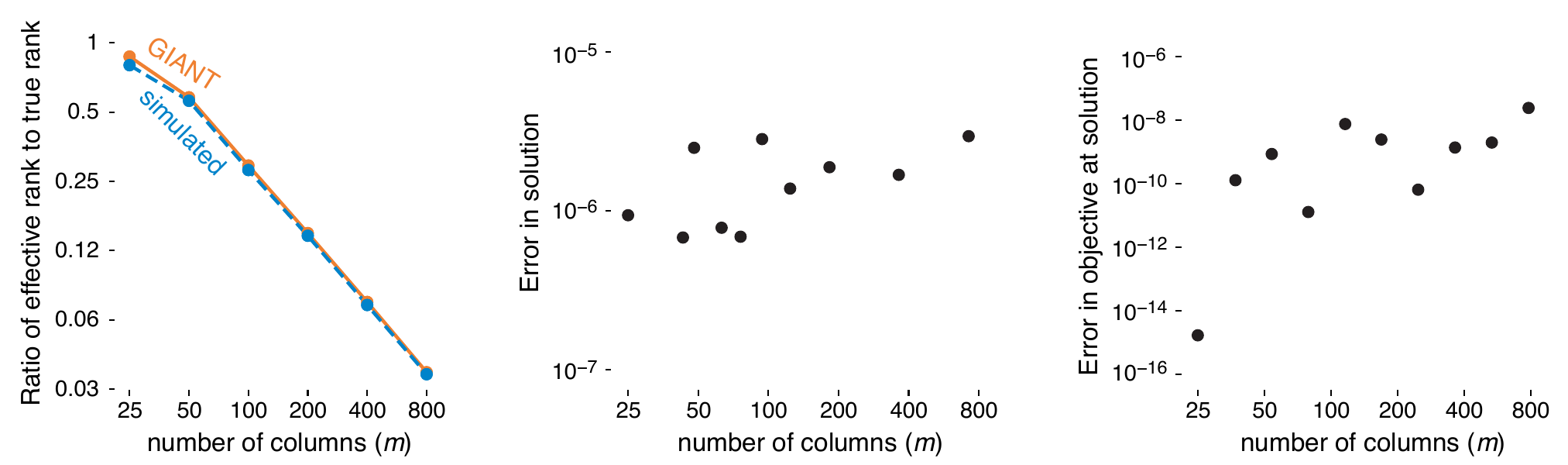}
\caption{Assessment of numeric rank of $L$ and its effect on solution
  accuracy. {\em Left panel:} The ratio of the effective rank $r$ (the
  rank estimated by {\tt pqrfact}, with relative tolerance of
  $10^{-10}$) to $m$, the number of columns. The ratios for the
  simulated data sets are averages from 10 simulations. {\em Middle
    panel:} $\ell_1$-norm of the differences in the solutions from the
  {\tt NN-SQP-A-F} and {\tt NN-SQP-A-QR} solvers applied to the GIANT
  data set. {\em Right panel:} Absolute difference in the objective
  values at these solutions. For all data sets used in these
  experiments, $n = \mbox{2,126,678}$.}
\label{fig:sparsity}
\end{center}
\end{figure}

The demonstrated benefits in using low-rank approximations are
explained by the fact that $r$, the effective rank of $L$, was small
relative to $m$ in our simulations. To check that this was not a
particular feature of our simulations, we applied the same SQP method
with RRQR ({\tt NN-SQP-A-QR}) to the GIANT data set. The ratio $r/m$ 
in the simulated and genetic data sets is nearly the same
(Figure~\ref{fig:sparsity}, left-hand panel).

We also assessed the impact of using low-rank approximations on the
quality of the solutions. For these comparisons, we used the RRQR
decomposition and the GIANT data set. In all settings of $m$ tested,
the error in the solutions was very small; the $\ell_1$-norm in the
difference between the solutions with exact and approximate $L$ was
close to $10^{-6}$ (Figure~\ref{fig:sparsity}, middle panel), and the
difference in the objectives was always less than $10^{-8}$ in
magnitude (Figure~\ref{fig:sparsity}, right-hand panel).

\begin{figure}[tb!]
\begin{center}
\includegraphics[width=5.25in]{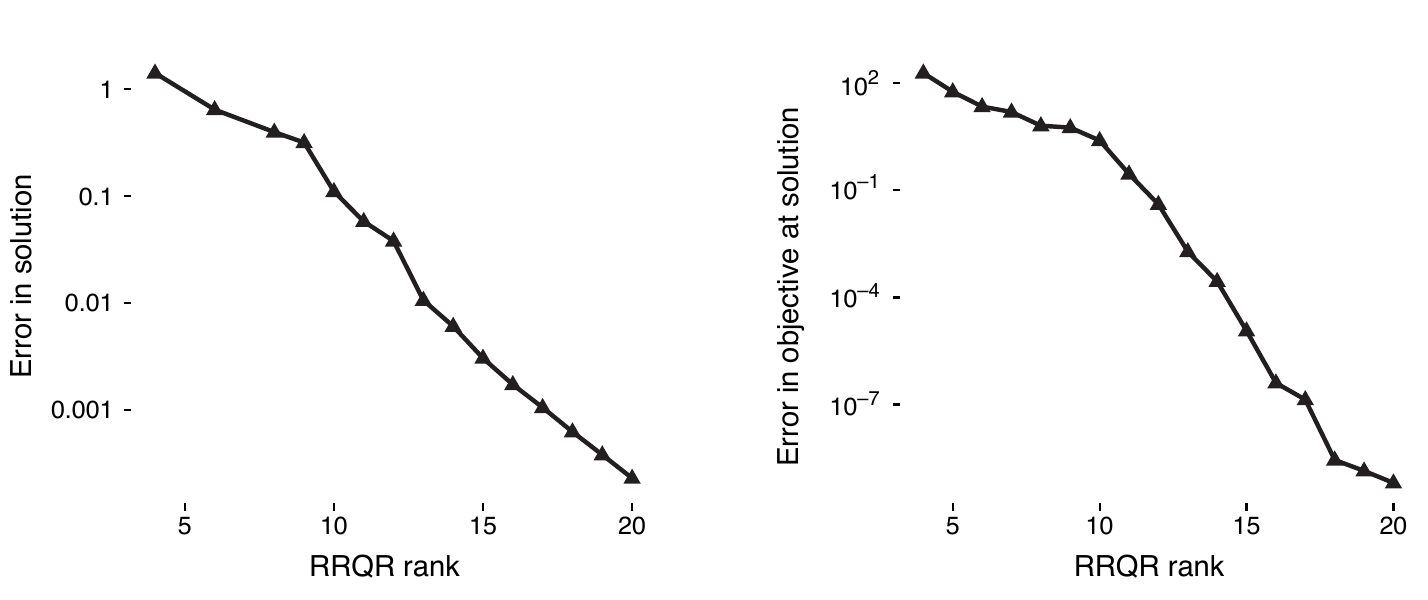}
\caption{Effect of QR rank ($r$) on accuracy of SQP solution. {\em
    Left panel:} $\ell_1$-norm of the difference in the solutions from
  the {\tt NN-SQP-A-F} and {\tt NN-SQP-A-QR} solvers applied to the
  same matrix $L$. {\em Right panel:} Absolute difference in the
  objective values at these solutions. All results are averaged over
  10 simulated data sets with $n = \mbox{100,000}$, $m = 200$ and $r$
  ranging from 4 to 20.}
\label{fig:lowrankaccuracy}
\end{center}
\end{figure}

To further understand how the RRQR approximation of $L$ affects
solution accuracy, we re-ran the SQP solver using QR approximations
with different ranks, rather than allow the ``rank revealing'' QR
algorithm to adapt the rank to the data.
Figure~\ref{fig:lowrankaccuracy} shows that the quality of the
approximate solution varies greatly depending on the rank of the QR
decomposition, and that the approximate solution gets very close to
the unapproximated solution as the rank is increased (presumably as it
approaches the ``true'' rank of $L$). These results illustrate the
importance of allowing the RRQR algorithm to adapt the low-rank
factorization to the data.

\subsubsection{Comparison of active set and IP solutions to quadratic
  subproblem}

\begin{figure}[tb!]
\begin{center}
\includegraphics[width=\textwidth]{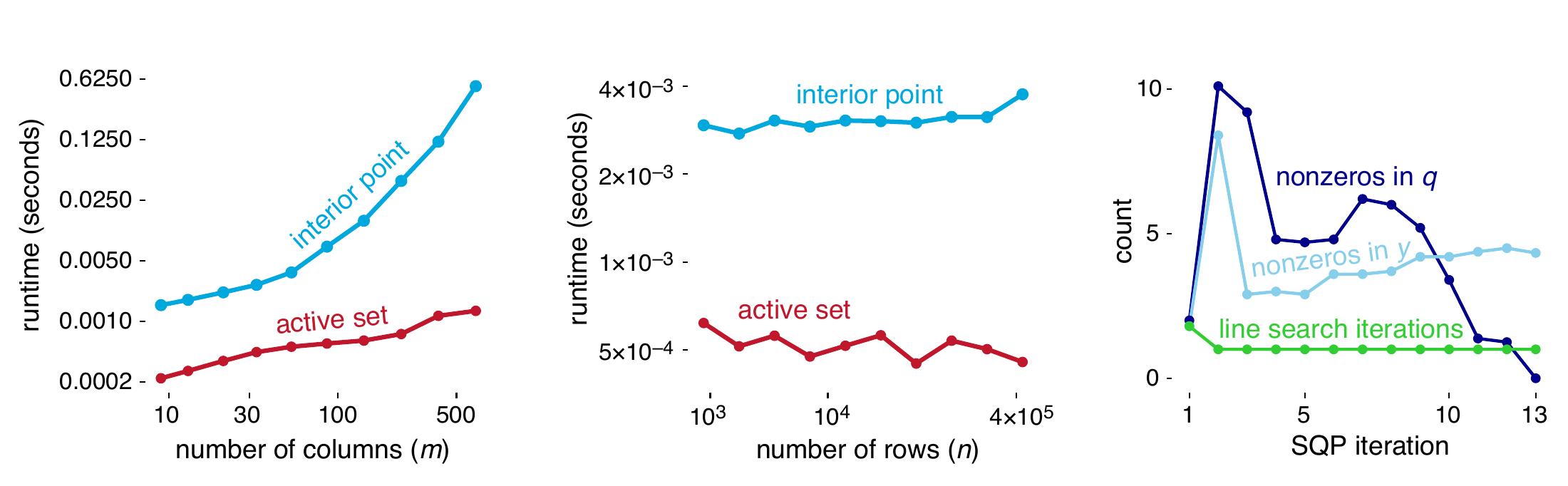}
\caption{Comparison of active set and interior point methods for
  solving the SQP quadratic subproblems. {\em Left panel:} Runtimes of
  the active set and IP ({\tt MOSEK}) methods as $m$ varies, with $n =
  10^5$. Runtimes are averaged over all subproblems solved, and over
  10 simulations. {\em Middle panel:} Runtimes of the IP and active
  set methods as $n$ varies, with $m = 40$. {\em Right panel:} Number
  of backtracking line searches, and the number of nonzero entries in
  $q$ and $y$ at each iteration of SQP, averaged over all SQP
  iterations and all 10 simulations. (Refer to
  eqs. \ref{prob:sqp_subproblem_reform} and
  \ref{prob:active_set_subprob} for interpreting $q$ and $y$.) }
\label{fig:ip_vs_activeset}
\end{center}
\end{figure}

In this set of experiments, we compared different approaches to
solving the quadratic subproblems inside the SQP algorithm: an active
set method (Section \ref{subsec:QP}) and an off-the-shelf IP method
({\tt MOSEK}); specifically, we compared {\tt NN-SQP-A-F} against {\tt
  NN-SQP-IP-F}. To assess effort, we recorded only the time spent
in solving the quadratic subproblems.

The left and middle plots in Figure~\ref{fig:ip_vs_activeset} show
that the active set method was consistently faster than the IP method
by a factor of roughly 5 or more, with the greatest speedups achieved
when $m$ and $n$ were large. For example, when $n = 10^5$ and $m
\approx 500$ in the left-hand plot, the active set solver was over 100
times faster than the IP method on average. The left-hand plot in
Figure~\ref{fig:ip_vs_activeset} shows that the complexity of the
active set solution to the quadratic subproblem grows linearly in $m$,
whereas the complexity of the IP solution grows quadratically. By
contrast, the average time required to solve the quadratic subproblems
does not depend on $n$ (see Figure~\ref{fig:ip_vs_activeset}, middle
panel), which could be explained by $n$ having little to no effect on
the number of degrees of freedom (sparsity) of the solution $x^{*}$.

We hypothesize that the active set method is faster because the quadratic
subproblem iterates and final solution are sparse (see the right-hand
plot in Figure~\ref{fig:ip_vs_activeset} for an illustration). Recall
that the reformulated problem \eqref{prob:main_problem} and the
quadratic subproblem \eqref{prob:sqp_subproblem} have a non-negative
constraint, which promotes sparsity.

Based on these results, we infer that the active set method
effectively exploits the sparsity of the solution to the quadratic
subproblem. We further note that poor conditioning of the Hessian may
favor the active set method because it tends to search over sparse
solutions where the reduced Hessian is better behaved.

In addition to the runtime improvements of the active set method,
another benefit is that it is able to use a good initial estimate when
available (``warm starting''), whereas this is difficult to achieve
with IP methods. Also, our active set implementation has the advantage
that it does not rely on a commercial solver. These qualitative
benefits may in fact be more important than performance improvements
considering that the fraction of effort spent on solving the quadratic
subproblems (either using the active set or IP methods) is relatively
small when $n \gg m$.

\subsubsection{Comparison of {\sc\bf mix-SQP} and KWDual}
\label{sec:mixsqp-vs-kwdual}

\begin{figure}[tb!]
\begin{center}
\includegraphics[width=4.25in]{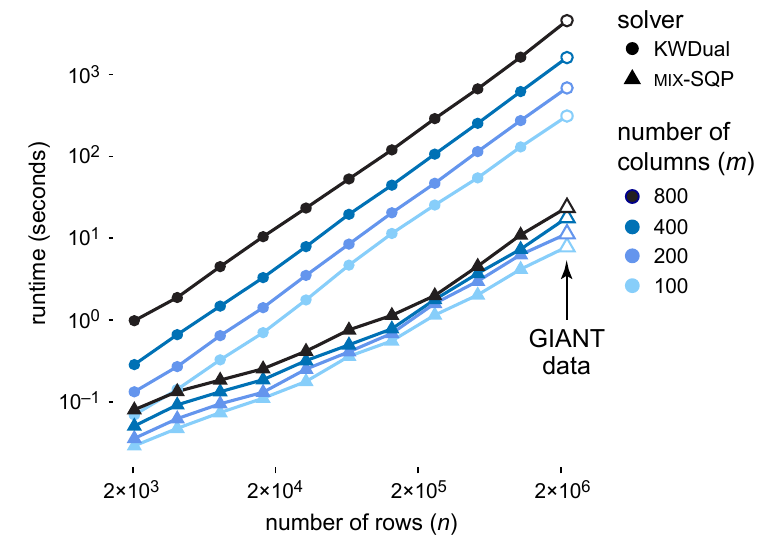}
\caption{Runtimes of \mixsqp{} and {\tt KWDual} (which uses {\tt
    MOSEK}) applied to simulated data sets with varying $n$ and $m$,
  and to the GIANT data set ($n = \mbox{2,126,678}$). All runtimes on
  simulated data sets were taken as averages over 10 separate
  simulations. Each timing on the GIANT data set is averaged over 10
  independent runs with the same $L$ matrix.}
\label{fig:real_data_summary}
\end{center}
\end{figure}

Based on the numerical results above, we concluded that when $n$ and
$m$ are large, and $n$ is larger than $m$, the fastest approach is
{\tt NN-SQP-A-QR}. We compared this approach, which we named
\mixsqp{}, against the {\tt KWDual} function from the R package {\tt
  REBayes} \citep{rebayes}, which is a state-of-the-art solver that
interfaces to the commercial software {\tt MOSEK} (this is {\tt
  D-IP-NA-F}). For fair comparison, all timings of {\tt KWDual} were
recorded in R so that communication overhead in passing variables
between Julia and R was not factored into the runtimes.

Although R often does not match the performance of Julia, an
interactive programming language that can achieve computational
performance comparable to C \citep{bezanson2012julia}, {\tt KWDual} is
fast because most of the computations are performed by {\tt MOSEK}, an
industry-grade solver made available as an architecture-optimized
dynamic library. Therefore, it is significant that our Julia
implementation consistently outperformed {\tt KWDual} in both the
simulated and genetic data sets; see Figure~\ref{fig:real_data_summary}.
For the largest data sets (e.g.,
$n \approx 10^6, m = 800$), \mixsqp{} was over 100 times faster than
{\tt KWDual}. Additionally, {\tt KWDual} runtimes increased more
rapidly with $m$ because {\tt KWDual} did not benefit from the
reduced-rank matrix operations.

\subsubsection{Assessing the potential for first-order optimization}
\label{sec:mixsqp-vs-first-order}

\begin{figure}[tb!]
\begin{center}
\includegraphics[width=6in]{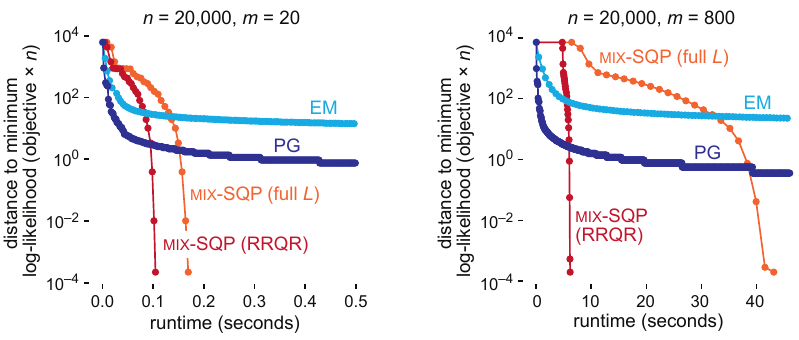}
\caption{Progress over time for the EM, projected gradient and
  \mixsqp{} methods on two simulated data sets with $n =
  \mbox{20,000}$ rows and $m = 20$ or $m = 800$ columns. The vertical
  axis shows the difference between the value of the log-likelihood,
  $f(x) \times n$, at the current iterate $x$, and the value of the
  log-likelihood at the best solution. Each dot corresponds to a
  single iteration of the algorithm's main loop.}
%
%
\label{fig:empgd}
\end{center}
\end{figure}

Our primary focus has been the development of fast methods for solving
\eqref{prob:mix_convex}, particularly when $n$ is large. For this
reason, we developed a method, \mixsqp{}, that makes best use of the
second-order information to improve convergence. However, it is
natural to ask whether \mixsqp{} is an efficient solution when $m$ is
large; the worst-case complexity is $O(m^3)$ since the active set step
requires the solution to a system of linear equations as large as $m
\times m$. (In practice, the complexity is often less that this worst
case because many of the co-ordinates are zero along the solution
path.) Here, we compare \mixsqp{} against two alternatives that avoid
the expense of solving an $m \times m$ linear system: a simple
projected gradient algorithm \citep{schmidt2009}, in which iterates
are projected onto the simplex using a fast projection algorithm
\citep{duchi2008efficient}; and EM \citep{dempster77}, which can be
viewed as a gradient-descent method \citep{xu-jordan-1996}.

The projected gradient method was implemented using Mark Schmidt's
MATLAB code.\footnote{{\tt minConF\_SPG.m} was retrieved from
  \url{https://www.cs.ubc.ca/spider/schmidtm/Software/minConf.html}.}
In brief, the projected gradient method is a basic steepest descent
algorithm with backtracking line search, in which the steepest descent
direction $-\nabla f(x)$ at iterate $x$ is replaced by
$\mathcal{P}_{\mathcal{S}}(x - \nabla f(x)) - x$, where
$\mathcal{P}_{\mathcal{S}}$ denotes the projection onto the feasible
set $\mathcal{S}^m$. As expected, we found that the gradient descent
steps were often poorly scaled, resulting in small step sizes. We kept
the default setting of 1 for the initial step size in the backtracking
line search. (The spectral gradient method, implemented in the same
MATLAB code, is supposed to improve the poor scaling of the steepest
descent directions, but we found that it was unstable for the problems
considered here.) The EM algorithm, which is very simple (see
Appendix \ref{sec:em}), was implemented in Julia.

To illustrate the convergence behaviour of the first-order methods, we
ran the approaches on two simulated data sets and examined the
improvement in the solution over time. Our results on a smaller
($\mbox{20,000} \times 20$) and a larger ($\mbox{20,000} \times 800$)
data set are shown in Figure~\ref{fig:empgd}. Both first-order methods
show a similar convergence pattern: initially, they progress rapidly
toward the solution, but this convergence slows considerably as they
approach the solution; for example, even after running EM and
projected gradient for 1,000 iterations on the $\mbox{20,000} \times
800$ data set, the solution remained at least 0.35 log-likelihood
units away from the solution obtained by \mixsqp{}. Among the two
first-order methods, the projected gradient method is clearly
the better option. Relative to the first-order methods, each iteration of
\mixsqp{} is very costly---initial iterations are especially slow
because the iterates contain few zeros at that stage---but \mixsqp{}
is able to maintain its rapid progress as the iterates approach the
solution. The benefit of the low-rank (RRQR) approximation in reducing
effort is particularly evident in the larger data set.

Although this small experiment is not intended to be a systematic
comparison of first-order vs. second-order approaches, it does provide
two useful insights. One, second-order information is crucial for
obtaining good search directions near the solution. Two,
gradient-descent methods are able to rapidly identify good approximate
solutions. These two points suggest that more efficient solutions for
large data sets, particularly data sets with large $m$, could be
achieved using a combination of first-order and second-order
approaches, or using quasi-Newton methods.

\subsubsection{Profiling adaptive shrinkage computations}

\begin{figure}[tb!]
\begin{center}
\includegraphics[width=\textwidth]{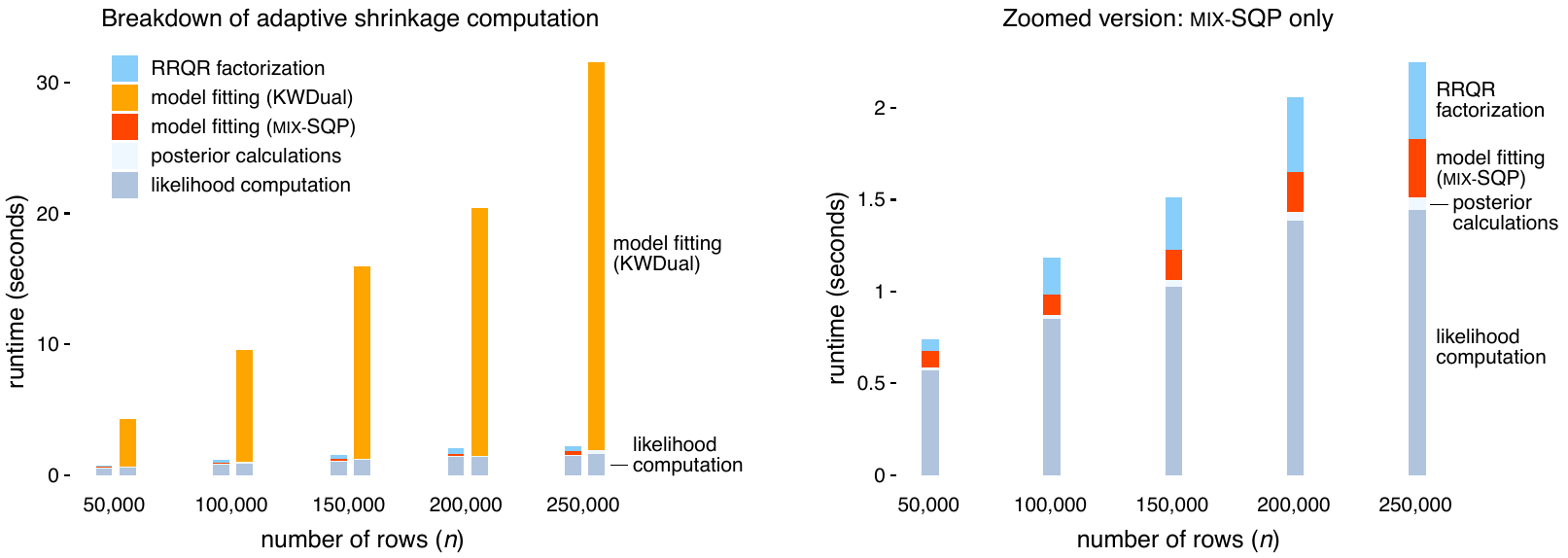}
\caption{Breakdown of computations for the ``adaptive shrinkage'' EB
  method \citep{ash}, in which the model-fitting step ({\em i.e.},
  maximum likelihood estimation of the mixture proportions) was
  implemented with either \mixsqp{} or {\tt KWDual} ({\tt MOSEK}). The
  adaptive shrinkage method was applied to simulated data sets with $m
  = 100$ and varying $n$. All runtimes were averaged over 10
  independent simulations. The right-hand panel is a zoomed-in version
  of the \mixsqp{} results shown in the left-hand plot.}
\label{fig:comp_time_breakdown}
\end{center}
\end{figure}

An initial motivation for this work was our interest in applying a
nonparametric EB method, ``adaptive shrinkage,'' to very large data
sets. This EB method involves three steps: (1) likelihood computation;
(2) maximum likelihood estimation of the mixture proportions; and (3)
posterior computation. When we began this work, the second step,
solved using {\tt MOSEK}, was the computational bottleneck of our R
implementation \citep{ashr}; see Figure~\ref{fig:comp_time_breakdown},
which reproduces the adaptive shrinkage computations in Julia (aside
from {\tt KWDual}). (To verify that this bottleneck was not greatly
impacted by the overhead of calling {\tt MOSEK} from R inside function
{\tt KWDual}, we also recorded runtime estimates outputted directly by
{\tt MOSEK}, stored in output {\tt MSK\_DINF\_OPTIMIZER\_TIME}. We
found that the overhead was at most 1.5 s, a small fraction of the
total model-fitting time under any setting shown in
Figure~\ref{fig:comp_time_breakdown}. Note all timings of {\tt KWDual}
called from Julia were recorded in R, not Julia.)

When we replaced {\tt KWDual} with \mixsqp{}, the model-fitting step
no longer dominated the computation time
(Figure~\ref{fig:comp_time_breakdown}). This result is remarkable
considering that the likelihood calculations for the scale mixtures of
Gaussians involve relatively simple probability density computations.

\section{Conclusions and potential extensions}
\label{sec:conclusions}

We have proposed a combination of optimization and linear algebra
techniques to accelerate maximum likelihood estimation of mixture
proportions. The benefits of our methods are particularly evident at
settings in which the number of mixture components, $m$, is moderate
(up to several hundred) and the number observations, $n$, is large. In
such settings, computing the Hessian is expensive---$O(nm^2)$
effort---much more so than Cholesky factorization of the Hessian,
which is $O(m^3)$. Based on this insight, we developed a sequential
quadratic programming approach that makes best use of the (expensive)
gradient and Hessian information, and minimizes the number of times it
is calculated. We also used linear algebra techniques, specifically
the RRQR factorization, to reduce the computational burden of gradient
and Hessian evaluations by exploiting the fact that the matrix $L$ often
has a (numerically) low rank. These linear algebra improvements were
possible by developing a customized SQP solver, in contrast to the use
of a commercial (black-box) optimizer such as MOSEK. Our SQP method also
benefits from the use of an active set algorithm to solve the
quadratic subproblem, which can take advantage of sparsity in the
solution vector. The overall result is that for problems with $n >
10^5$, \mixsqp{} can achieve a 100-fold speedup over {\tt KWDual},
which applies the commercial {\tt MOSEK} interior point solver to a
dual formulation of the problem.

To further reduce the computational effort of optimization in
nonparametric EB methods such as \cite{koenker2014convex} and
\cite{ash}, {\em quasi-Newton methods} may be fruitful
\citep{nocedal2006sequential}. Quasi-Newton methods, including the
most popular version, BFGS, approximate $H$ by means of a secant
update that employs derivative information without ever computing the
Hessian. While such methods may take many more iterations compared
with exact Hessian methods, their iterations are $m$ times
cheaper---consider that evaluating the gradient
\eqref{eqn:grad_hess_qr} is roughly $m$ times cheaper than the Hessian
when $n \gg m$---and, under mild conditions, quasi-Newton methods
exhibit the fast superlinear convergence of Newton methods when
sufficiently close to the solution \citep{nocedal2006sequential}.

Since $n$ is the dominant component of the computational complexity in
the problem settings we explored---the gradient and Hessian
calculations scale linearly with $n$---another promising direction is
the use of {\em stochastic approximation} or {\em online learning}
methods, which can often achieve good solutions using approximate
gradient (and Hessian) calculations that do not scale with $n$
\citep{bottou, robbins-monro}. Among first-order methods,
\textit{stochastic gradient descent} may allow us to avoid linear
per-iteration cost in $n$. In the Newton setting, one could explore
stochastic quasi-Newton \citep{byrd2016} or LiSSA \citep{agarwal2016}
methods.

As we briefly mentioned in the results above, an appealing feature of
SQP approaches is that they can easily be warm started. This is much
more difficult for interior point methods \citep{potra2000interior}.
``Warm starting'' refers to sequential iterates of a problem becoming
sufficiently similar that information about the subproblems that is
normally difficult to compute from scratch (``cold'') can be reused as
an initial estimate of the solution. The same idea also applies to
solving the quadratic subproblems. Since, under general assumptions,
the active set settles to its optimal selection before convergence,
this suggests that the optimal working set $\mathcal{W}$ for
subproblem $\mathcal{P}$ will often provide a good initial guess for
the optimal working set $\mathcal{W}^{\star}$ for similar subproblem
$\mathcal{P}^{\star}$.

\subsection*{Acknowledgments}

This material was based upon work supported by the U.S. Department of Energy, Office of Science, Office of Advanced Scientific Computing Research (ASCR) under Contract DE-AC02-06CH11347. We acknowledge partial NSF funding through awards FP061151-01-PR and CNS-1545046 to MA, and support from NIH grant HG002585 and a grant from the Gordon
and Betty Moore Foundation to MS. We thank the staff of the
University of Chicago Research Computing Center for providing
high-performance computing resources used to implement some of
the numerical experiments. We thank Joe Marcus for his help in
processing the GIANT data, and other members of the Stephens
lab for feedback on the methods and software.
        


\bibliographystyle{agsm}
\bibliography{mixsqp}

\vspace{-0.15cm}
\begin{flushright}
	\scriptsize \framebox{\parbox{2.5in}{Government License: The
            submitted manuscript has been created by UChicago Argonne,
            LLC, Operator of Argonne National Laboratory
            (``Argonne'').  Argonne, a U.S. Department of Energy
            Office of Science laboratory, is operated under Contract
            No. DE-AC02-06CH11357.  The U.S. Government retains for
            itself, and others acting on its behalf, a paid-up
            nonexclusive, irrevocable worldwide license in said
            article to reproduce, prepare derivative works, distribute
            copies to the public, and perform publicly and display
            publicly, by or on behalf of the Government. The
            Department of Energy will provide public access to these
            results of federally sponsored research in accordance with
            the DOE Public Access
            Plan. http://energy.gov/downloads/doe-public-access-plan. }}
        \normalsize
\end{flushright}

\clearpage

\setcounter{page}{1}

\appendix

\section{EM for maximum-likelihood estimation of mixture
  proportions}
\label{sec:em}

Here we derive the EM algorithm for solving
\eqref{prob:mix_convex}. The objective $f(x)$ can be recovered as the
log-likelihood (divided by $n$) for $z_1, \ldots, z_n$ drawn {\em
  i.i.d.} from a finite mixture, in which the $x_k$'s specify the
mixture weights:
\begin{equation*}
p(z_j \,|\, x) \sim \sum_{k=1}^m x_k g_k(z_k), \quad
\mbox{for $j = 1, \ldots, n$.}
\end{equation*}
The mixture model is equivalently formulated as
\begin{align*}
p(\gamma_j = k \,|\, x)   &= x_k \\ 
p(z_j \,|\, \gamma_j = k) &= g_k(z_j), \quad
  \mbox{for $j = 1, \ldots, n$,} 
\end{align*}
in which we have introduced latent indicator variables $\gamma_1,
\ldots, \gamma_n$, with $\gamma_j \in \{1, \ldots, m\}$.

Under this augmented model, the expected complete log-likelihood is
\begin{align*}
E[\log p(z \,|\, x, \gamma)] &=
\sum_{j=1}^n
E\big[\log\big\{ p(z_j\,|\,\gamma_j) \, p(\gamma_j \,|\, x)\big\}\big] \\
&= \sum_{j=1}^n \sum_{k=1}^K \phi_{jk} \log(L_{jk} x_k),
\end{align*}
where $\phi_{jk}$ denotes the posterior probability $p(\gamma_j = k
\,|\, x, z_j)$, and $L_{jk} \triangleq g_k(z_j)$. From this expression
for the expected complete log-likelihood, the M-step update for the
mixture weights works out to
\begin{equation}
x_k = \frac{1}{n} \sum_{j=1}^n \phi_{jk},
\label{eq:M-step}
\end{equation}
and the E-step consists of computing the posterior probabilities,
\begin{equation}
\phi_{jk} = \frac{L_{jk} x_k}{\sum_{k' = 1}^m L_{jk'} x_{k'}}.
\label{eq:E-step}
\end{equation}

In summary, the EM algorithm for solving \eqref{prob:mix_convex} is
easy to explain and implement: it iteratively updates the posterior
probabilities for the current estimate of the mixture weights,
following eq.~\ref{eq:E-step} (this is E-step), then updates the
mixture weights according to eq.~\ref{eq:M-step} (this is the M-step),
until some stopping criterion is met or until some upper limit on the
number of iterations is reached.

\section{Proofs}
\label{sec:proofs}

Here we provide proofs for Proposition~\ref{thm:equivalence} and
Proposition~\ref{corr:special_equiv}.

\subsection{Proof of Proposition \ref*{thm:equivalence}}

Because of the monotonicity and unboundedness of the objective over
the positive orthant, $\mathbb{R}^m$, the solution to
\eqref{prob:simplex} is preserved if we relax the simplex constraint
$\mathcal{S}^m = \{x : \mathds{1}^Tx = 1, x \succeq 0\}$ to a set of
linear inequality constraints, $\{x : \mathds{1}^Tx \leq 1, x \succeq
0\}$.  Slater's condition \citep{boyd} is trivially satisfied for both
formulations of the simplex constraints, and the feasible set is
compact. The solution then satisfies the KKT optimality conditions;
{\em i.e.}, at the solution $x^{*}$ there exists a $\lambda^{*} >
0$ and a $\mu^{*} \succeq 0$ such that
\begin{equation}
\begin{aligned}
\nabla \phi(x^{*}) + \lambda^{*} \mathds{1} - \mu^{*} &= 0 \\
(\mu^{*})^T x^{*} &= 0 \\
\mathds{1}^T x ^* &= 1.
\end{aligned}
\label{eq:KKT}
\end{equation}
We therefore conclude that solving \eqref{prob:simplex} is equivalent
to solving
\begin{equation}
\label{eqn:lag_relax}
\underset{x \,\in\, \mathbb{R}_{+}^{m}}{\textrm{minimize}} \;
\phi(x) + \lambda^{*} \big({\textstyle \sum_{k=1}^m x_k - 1} \big).
\end{equation}
We claim that for any $\lambda > 0$, the solution to the Lagrangian
relaxation of the problem,
\begin{equation*}
x^{*}(\lambda) = \argmin_{x \in \mathbb{R}_{+}^m}
\phi(x) + \textstyle \lambda \sum_{k=1}^m x_k,
\end{equation*}
is the same as solution to the original problem up to a constant of
proportionality, and the proportionality constant is $\lambda /
\lambda^{*}$. So long as $\lambda^{*} > 0$, we have that
\begin{align*}
x^{*}(\lambda) &= \argmin_{x \,\in\, \mathbb{R}_{+}^m}
\textstyle \big\{ \phi(x) + \lambda \sum_{k=1}^m x_k \big\} \\
&= \argmin_{x \,\in\, \mathbb{R}_{+}^m} \textstyle \big\{
\phi(\frac{\lambda^{\;}}{\lambda^{*}} x) + \lambda^{*}
\sum_{k=1}^m \frac{\lambda^{\;}}{\lambda^{*}} x_k \big\} \\
&= \frac{\lambda^{*}}{\lambda^{\;}} \times
\argmin_{x' \,\in\, \mathbb{R}_+^m} \textstyle \big\{ \phi(x') +
\lambda^{*} \sum_{k=1}^m x_k' \big\} \\
&= \textstyle \frac{\lambda^{*}}{\lambda^{\;}} x^*(\lambda^*).
\end{align*}
The second equality follows from the scale invariance assumption on
$\phi(x)$.

Note that $\lambda^{*} = 0$ cannot hold at a solution to
\eqref{prob:simplex}. Suppose that $\lambda^{*} = 0$. Then the point
$x^{*}$ must satisfy the KKT conditions of the problem in which the
equality constraint $\mathds{1}^T x = 1$ is removed, and it must then
be a solution of that problem. Since the objective function decreases
as we scale up $x$, such problems clearly are unbounded below and thus
cannot have an optimal solution (as scaling $x$ up keeps decreasing
the function value while preserving nonnegativity of entries in $x$).
Since the solution of \eqref{prob:simplex} must satisfy $\mathds{1}^Tx
= 1$, the conclusion follows.

\subsection{Proof of Proposition \ref*{corr:special_equiv}}

The proof follows from the KKT optimality conditions
\eqref{eq:KKT}. Premultiplying the first set of equations by
$(x^{\star})^T$, we have that
\begin{equation*}
(x^{*})^T \nabla f(x^{*}) + \lambda^{*} = 0.
\end{equation*}
The gradient of the objective in \eqref{prob:mix_convex} is $\nabla f(x)
= -L^Td/n$, where $d$ is a vector of length $n$ with entries $d_j =
1/(Lx)_j$. Inserting this expression for the gradient into the above
identity yields $\lambda^{*} = 1$.

\section{Implementation of the active set method}

At each iteration of \mixsqp{} (Algorithm~\ref{alg:sqp}), an active
set method is used to compute a search direction, $p^{(t)}$. The
active set method is given in Algorithm~\ref{alg:subprob}. It is
adapted from Algorithm~16.3 of \cite{nocedal2006sequential}. Note that
additional logic is needed to handle boundary conditions, such as the
case when the working set is empty, and when the working set contains
all co-ordinates except one.

\begin{algorithm}[tb]
\SetKwInOut{Input}{Inputs}
\SetKwInOut{Output}{Output}
\Input{gradient, $g \in \mathbb{R}^m$; Hessian, $H \in
  \mathbb{R}^{m \times m}$; initial working set,
  $\mathcal{W}^{(0)} \subseteq \{1,\ldots,m\}$;
  convergence tolerance, $\epsilon \geq 0$.}
\Output{$y^{(l)} \in \mathbb{R}^m$, an estimate of the solution to
  \eqref{prob:sqp_subproblem_reform}.}
$k \leftarrow m - |W^{(0)}|$; \\
Set $y_i^{(0)} \leftarrow 0, \forall i \in \mathcal{W}^{(0)}$; \\
Set $y_i^{(0)} \leftarrow 1/k, \forall i \notin \mathcal{W}^{(0)}$; \\
\For {l = {\rm 0, 1, 2, \ldots}} {
  $b \leftarrow Hy^{(l)} + 2g + \mathds{1}_m$; \,
    (see eq. \ref{prob:active_set_subprob}) \\
  $q^{(l)} \leftarrow \argmin_q q^THq/2 + q^Tb$ s.t.
  $q_i = 0, \forall i \in \mathcal{W}^{(l)}$; \,
  (see eq. \ref{prob:active_set_subprob}) \\
  $\alpha_l \leftarrow 1$; \\
  \If{$\max_i |q_i^{(l)}| \leq 0$}{
  \If{$\min_{i\,\in\,\mathcal{W}^{(l)}} b_i \geq -\epsilon$ }{
    {\bf stop}; \, (all Lagrange multipliers in working set are positive)
  }
  {
  $j \leftarrow \argmin_{i\,\in\,\mathcal{W}^{(l)}} b_i$; \\
  $\mathcal{W}^{(l+1)} \leftarrow \mathcal{W}^{(l)} \setminus \{j\}$;
    \, (remove smallest multiplier from active set)
  }
  }
  \Else{
    $j \leftarrow \argmin_{i \,\notin\, \mathcal{W}^{(l)}, \, q_i^{(l)} < 0}
      -y_i^{(l)}/q_i^{(l)}$; \\
    $\alpha_l \leftarrow -y_j^{(l)}/q_j^{(l)}$;
      (find largest step size retaining feasibility) \\
    \If {$\alpha_l < 1$} {
      $\mathcal{W}^{(l+1)} \leftarrow \mathcal{W}^{(l)} \cup \{j\}$; \,
      (add blocking constraint to working set)
    }
  }
  $y^{(l+1)} \leftarrow y^{(l)} + \alpha_l q^{(l)} $\;
}
\caption{{\sc mix-active-set}: active set method method to compute a
  search direction for \mixsqp{}.\label{alg:subprob}}
\end{algorithm}

\section{GIANT data processing details}

We retrieved file {\tt
  GIANT\_HEIGHT\_Wood\_et\_al\_2014\_publicrelease\_HapMapCeuFreq.txt.gz}
from the GIANT Project Wiki
(\url{http://portals.broadinstitute.org/collaboration/giant}). The
original tab-delimited text file contained summary
statistics---regression coefficient estimates $z_j$ and their standard
errors $s_j$ (columns ``b'' and ``SE'' in the text file)---for
2,550,858 SNPs $j$ on chromosomes 1--22 and chromosome X. These
summary statistics were computed from a meta-analysis of 79
genome-wide association studies of human height; see \cite{giant} for
details about the studies and meta-analysis methods used. We filtered
out 39,812 SNPs that were not identified in Phase 3 of the 1000
Genomes Project \citep{1kg}, an additional 384,254 SNPs where the
coding strand was ambiguous, and 114 more SNPs with alleles that did
not match the 1000 Genomes data, for a final data set containing $n =
\mbox{2,126,678}$ SNPs. (Note that the signs of the $z_j$ estimates
were flipped when necessary to align with the 1000 Genomes SNP
genotype encodings, although this should have had no effect on our
results since the prior is a mixture of zero-centered normals.) The
processed GIANT data are included in the accompanying source code
repository.

\end{document}